# Influence of Environmentally Affected Hole Transport Layers on Spatial Homogeneity and Charge Transport Dynamics of Organic Solar Cells.

Huei-Ting Chien[1*], Florian Pilat[1], Thomas Griesser[2], Harald Fitzek[3], Peter Poelt[3,4], Bettina Friedel[1,5]

[1]Institute of Solid State Physics, Graz University of Technology, Austria;

[2]Chair of Chemistry of Polymeric Materials, University of Leoben, Austria;

[3]Institute of Electron Microscopy and Nanoanalysis, Graz University of Technology, Austria;

[4] Graz Centre for Electron Microscopy, Austria

[5]illwerke vkw Endowed Professorship for Energy Efficiency, Energy Research Center, Vorarlberg University of Applied Sciences, Austria

*Email: hchien@tugraz.at

ABSTRACT

After the efficiency of organic photovoltaic (OPV) cells achieved more than 10%, the control of stability and degradation mechanisms of solar cells became a prominent task. The improvement of device efficiency due to incorporation of a hole-transport layer (HTL) in bulk-heterojunction solar cells has been extensively reported. However, the most widely used HTL material, poly(3,4-ethylenedioxythiophene):poly(styrenesulfonate) (PEDOT:PSS) is frequently suspected to be the dominating source for device's instability under environmental conditions. Thereby effects like photooxidation and electrode corrosion are often reported to shorten device lifetime. However, often in environmental device studies, the source of degradation, whether being from the HTL, the active layer or the metal cathode are rather difficult to distinguish, because the external diffusion of oxygen and water affects all components. In this study, different HTLs, namely prepared from traditional PEDOT:PSS and also two types of molybdenum trioxide ($MoO_3$), are exposed to different environments such as oxygen, light or humidity, prior to device finalization under inert conditions. This allows investigating any effects within the HTL and from reactions at its interface to the indium-tin-oxide electrode or the active layer. The surface and bulk chemistry of the exposed HTL has been monitored and discussed in context to the observed device physics, dynamic charge transport and spatial performance homogeneity of the according OPV device. The results show that merely humidity-exposure of the HTL leads to decreased device performance for PEDOT:PSS, but also for one type of the tested $MoO_3$. The losses are related to the amount of absorbed water in the HTL, inducing loss of active area in terms of interfacial contact. The device with PEDOT:PSS HTL after humid air exposure showed seriously decreased photocurrent by micro-delamination of swelling/shrinkage of the hygroscopic layer. The colloidal $MoO_3$ with water-based precursor solution presents slight decay of solar cell performance, also here caused by swelling/shrinking reaction, but by a combination of in-plane particle contact and resistance scaling with particle expansion. However, the device with quasi-continuous and alcohol-based $MoO_3$ showed unharmed stable electrical performance.

INTRODUCTION

The research on organic photovoltaics (OPV) reached remarkable growth due to their potential for realization of flexible, large-area and light-weight devices, categorized by environmental friendliness, low thermal budget and low-cost production e.g. via roll-to-roll printing.[1–4]. The power conversion efficiency (PCE) of OPVs has been greatly improved by introducing the bulk-heterojunction (BHJ) concept by the combination of organic donor and acceptor on the nanoscale. [1,5] The most prominent system here is the donor-acceptor couple of the conjugated polymer poly(3-hexylthiophene) (P3HT) and the fullerene derivative phenyl-$C_{61}$-butyric acid methyl ester (PCBM). [6–7] Best systems currently reach up to 12% efficiency.[8-9] Typical BHJ OPVs require a selective hole-transport layer (HTL) to block electrons and enhance hole-transfer to the anode, further to stabilize the anode's work function and reducing its surface roughness. [10-11] The most commonly used material for the anode interlayer is poly(3,4-ethylenedioxythiophene):poly(styrenesulfonate) (PEDOT:PSS) due to its benefits such as high conductivity, high transparency and good solution processibility[12-13]. However, beside high efficiencies and cost-effective processing, also long device lifetimes are desirable for OPVs. [14] Plenty of research has been focusing on environmentally induced mechanisms leading to device degradation, based on various approaches and techniques. [15–20] This is not exclusively caused by environments like air, light and humidity, even solvent vapors which are present during fabrication or even testing, can affect the device performance considerably.[71] But exactly the long-term stability is a concern when using PEDOT:PSS, since the strongly hygroscopic polysulfonic acid dissociates under the influence of humidity, suspected to induce corrosion in the device. [14-21] To overcome this problem, high-work function transition metal oxides, such as $NiO_x$ [22–23], $WO_3$ [24] and $MoO_3$ [25-27] have been introduced as alternatives to PEDOT:PSS in OPVs. Still, it is not clear if those materials are entirely inert against environmental influence.

Spatially resolved methods such as conductive atomic force microscopy [28], near-field scanning optical microscopy [29], lock-in thermography [30], photo- and electroluminescence [31] and light-beam induced current (LBIC) [27,31] have been applied as tools to characterize surface and device performance of OPVs. In particular, laser beam induced current (LBIC) has been widely used to investigate degradation and efficiency losses by spatial homogeneity across the entire photoactive pixel area. [20,32-34] Further, transient photocurrents (TPC) have been used to gain knowledge about the dynamics of charge separation, determine mobilities, to describe the build-up of trapped charges and the dynamics of recombination of OPVs. [35-41]

Aim of this work is to distinguish the environmental effects solely induced by exposition of the HTL, without the impact of a total device exposure. Therefore three different types of solution processed HTLs have been compared: (1) deposited from commercial PEDOT: PSS ink, (2) a quasi-continuous $MoO_3$ nanolayer obtained from deposition of an acetylacetonate precursor in alcohol and hydrolysation in air and (3) a $MoO_3$ layer of colloidal particles obtained from hydrolysis and condensation of an molybdate precursor in acidic aqueous solution. The two different precipitation methods for $MoO_3$ were chosen to see eventual effects of an acid presence. The as-prepared HTLs on indium-tin-oxide (ITO) glass substrates are exposed to dry air, humid air in the dark or with illumination, before further processing of the ITO/HTL/P3HT:PCBM/LiF/Al device stack in inert atmosphere. Surface and local bulk chemistry of the exposed HTLs have been investigated by X-ray photoelectron spectroscopy (XPS) and Raman microscopy, respectively. Finalized devices are characterized regarding their general device performance and scanned regarding their photocurrent homogeneity and local device characteristics by LBIC. The dynamic behavior of charge transport in the devices was studied by TPC. Altogether, these results allow conclusions on device changes/degradation solely connected to chemical/photochemical reactions of the HTL itself or at the closest interfaces in consequence.

EXPERIMENTAL SECTION

**Materials**

P3HT was supplied by Rieke Metals Inc. (MW 50–70 kg mol$^{-1}$, regioregularity 91–94%). PC$_{60}$BM was purchased from Nano-C Inc. (99.5% purity). Anhydrous chlorobenzene (99.8%) as solvent for the organic semiconductors, was purchased from Acros Organics. The aqueous formulation of PEDOT: PSS ink was purchased from Heraeus Deutschland GmbH & Co. KG (Clevios P Jet (OLED)). Ammonium molybdate ((NH4)$_2$MoO$_4$) (≥99.98%), hydrochloric acid (HCl) (≥37%) and bis(acetylacetonato)dioxomolybdenum(VI) (MoO$_2$(acac)$_2$) were purchased from Sigma-Aldrich. Isopropanol (99.9%) was purchased from VWR International LLC. All substances were used as received. ITO substrates (20 Ω /square, Ossila) were cleaned by sonication in acetone and isopropanol, followed by O$_2$ plasma etching (100 W for 30 min) briefly before further processing.

**Preparation of the MoO$_3$ and PEDOT:PSS HTLs**

Two different formulations were used to generate the MoO$_3$ films. The quasi-continuous films, consisting of a nanosized network with particle mean diameter of 6 nm, [27] are referred to in the following as *MoO$_3$-1*. For their preparation, following a procedure reported by K. Zilberberg et al., [25] MoO$_2$(acac)$_2$ was dissolved in isopropanol to form a 0.5% (w/v) precursor solution. The colloidal films, comprising a layer of separated particles (10-30 nm), [27] are referred to as *MoO$_3$-2* in this study and were prepared as reported by Liu et al. [26]. Here, (NH$_4$)$_2$MoO$_4$ was dissolved in distilled water to form a 0.005 mol/L solution. Then 2 mol/L aqueous HCl solution as catalyst was added drop-wise under stirring until the pH value of the solution was between 1 and 1.5. Particles of MoO$_3$ start forming already in solution by hydrolysis and condensation, thus generating a suspension. All three formulations, PEDOT:PSS, MoO$_3$-1 and MoO$_3$-2, were filtered by 0.22 μm PVDF membrane filters (Sigma Aldrich) prior to deposition and spin-coated onto ITO substrates at 4000 rpm for PEDOT: PSS and 3000 rpm for MoO$_3$ for 40s, respectively. The layers of MoO$_3$-1 and MoO$_3$-2 were kept in ambient air at room temperature for 1 h for hydrolysis of the precursor. Finally, films were annealed at 160 °C for 20 min to dry and promote oxide formation. The PEDOT: PSS layers were merely heated to 160 °C for 20 min under Argon (Ar) flow for removal of water. The film thickness of PEDOT:PSS is around 40 nm, MoO$_3$-1 and MoO$_3$-2 are both around 10 nm. The as-prepared HTLs were either used immediately for device fabrication (referred to as *\*/Fresh*) or underwent selected exposure before. For this purpose films were submitted to dry air with <20% relative humidity (*\*/Air*) or humid air with >80% relative humidity (*\*/H$_2$O*) without any illumination, or to dry air under white light illumination (tungsten halogen) at 10% equivalent of AM1.5G (*\*/Air & light*), respectively. For each condition, the temperature was 25°C and the duration of exposure was 18 hours. Devices were fabricated immediately after exposure. Samples for XPS investigation were prepared with the same procedure on unpatterned ITO substrates. Different to the standard process, samples for Raman spectroscopy were prepared on ITO substrates for PEDOT:PSS and on Si substrates for MoO$_3$-1 and MoO$_3$-2 by drop casting method to obtain thicker layers (~80nm) to increase the weak signal intensity.

**Device fabrication**

Standard solar cells with P3HT:PCBM photoactive layer according to the architecture shown in Figure 1 were assembled in inert gas atmosphere, using named HTLs of *MoO$_3$-1*, *MoO$_3$-2* and *PEDOT: PSS* directly after different environments exposures. The active layer was applied on the according ITO/HTL substrate by spin-coating from a solution of P3HT and PCBM (1:1 weight ratio, 18 mg/mL each) in 70°C chlorobenzene at 1500 rpm for 60 s, followed by annealing at 120° C for 5 min. The film thickness obtained is around 150 nm. The cathode was thermally evaporated as a bilayer of LiF (2nm) and Al (100 nm). The nominal active area of the devices is 6 mm$^2$ (pixel size 1.5 mm x 4.0 mm).

**Characterization**

The surface chemistry of HTLs was characterized by XPS and local bulk chemistry by Raman spectroscopy. The XPS (Thermo Fischer Scientific, Waltham, MA USA) is equipped with a monochromatic Al Kα X-ray source (1486.6 eV). All scans were performed at room temperature. The peaks were fitted using a Gaussian/Lorentzian mixed function employing Shirley background correction (Software XPSPEAK41). Raman measurements were performed with a Horiba Jobin Yvon LabRam 800 HR spectrometer equipped with a 1024 x 256 CCD (Peltier-cooled) and an Olympus BX41 microscope. All scans were done using a laser wavelength of 532 nm (50 mW), an x50 Olympus LMPlanFLN (N.A. 0.5) objective and a 300 l/mm grating (spectral resolution of 3 cm$^{-1}$). Spatial Raman spectroscopy characteristics for point-to-point analysis were taken in line-scans along 100 μm scanning area with 2 μm step size for PEDOT:PSS, and with 4 μm for MoO$_3$-1 and MoO$_3$-2, respectively. In order to avoid damage to the sample, the laser intensity was reduced to 10 % for the measurements of PEDOT:PSS and MoO$_3$-2 and to 1 % for the measurements of MoO$_3$-1. The photocurrent density–voltage (J–V) characteristics of the OPV cells were recorded using a computer-controlled Keithley 2636A source meter and AM1.5G illumination (100 mW/cm$^2$) from a solar simulator (Model 10500,

ABET Technologies, rated ABB). Spatial photocurrent distribution was determined by laser beam induced current (LBIC) method, scanning with a computer-controlled nano-manipulator-driven xy-stage (Kleindiek Nanotechnik, NanoControl NC-2-3) and excitation with a 532 nm laser (<5mW) with a focused spot-size of ≈2 µm. Spatial photocurrent mappings were scanned with 40 µm step size across the entire photoactive pixel area of 4.0 mm x 1.5 mm (Fig. 1). Local photocurrent-voltage characteristics for point-to-point analysis were taken in line-scans along the long side of each pixel, scanning from the left (0 mm) to the right (4 mm) with 160 µm step size. For transient photocurrent measurements, a 530 nm green LED (Thorlabs, M530L2) was used as the light source and a 300 µs square pulse from a function generator (HAMEG HMF 2550) at 1 kHz was used to power the LED. The photocurrent of the devices was measured by an oscilloscope (HAMEG HMO 3002 Series) with input impedance of 50 Ω. For voltage-dependent photocurrent transients, devices were connected to a computer-controlled Keithley 2636A source meter operating in DC mode.

RESULTS AND DISCUSSION

To investigate how the environmental conditions such as air, illumination and humidity affect the HTLs and in consequence lead to degradation of the OPV devices, solar photocurrent density-voltage (J-V) characteristics were recorded, shown in Figure 2. In every case the "fresh" condition acts as blind sample in this study, to account for environment-independent relaxation, aging or measurement-related changes of the materials. Figure 2(a) shows the J-V-characteristics of devices with exposed PEDOT:PSS HTLs. It can be seen that **PEDOT:PSS/Air** and **PEDOT:PSS/Air & light** present the best performance of the 4 conditions, with the highest rectangularity indicated by fill factors (FF) of 0.49 and 0.50, respectively. **PEDOT:PSS/Fresh** on the other hand, shows equal short-circuit current ($J_{SC}$) of around -8.6 mA/cm$^2$ and open-circuit voltage ($V_{OC}$) of around 0.54, but considerably lower fill factor of only 0.44. However, the by far lowest performance is shown by **PEDOT:PSS/H₂O**, which exhibits lower rectangularity with FF of 0.39 and additionally greatly reduced photocurrent of $J_{SC}$ of -6.4 mA/cm², but stable $V_{OC}$ of 0.55. The behaviour of the fresh sample compared to the air- and light-exposed ones can be explained by reduced interfacial charge transport efficiency between PEDOT:PSS and neighboring layers and an associated higher series resistance of the device, indicated by low FF and the change in slope around the $V_{OC}$, respectively. The reason for the lack in interfacial transfer is the non-equilibrium condition the PEDOT:PSS layer is in when it is processed into a device right away. In that case, a certain degree of aging is necessary for relaxation of the polymer and its surface groups to make better contact to the ITO electrode and the active layer. The low FF accompanied by reduced photocurrent in **PEDOT:PSS/H₂O** on the other hand, is a sign for loss of active area, in the sense of ITO corrosion producing insulating islands or local (photo)degradation of the adjacent active area facilitated by the absorbed water diffusing from the exposed PEDOT:PSS layer.[42-44] Figure 2(b) shows the solar J-V curve of devices with MoO₃-1 HTL after different environmental exposures. All four samples, **MoO₃-1/Fresh**, **MoO₃-1/Air**, **MoO₃-1/Air & light** and **MoO₃-1/H₂O** show identical unaltered performance irrespective of the exposure of the HTL, with $J_{SC}$ of about -9.5 mA/cm², a $V_{OC}$ of 0.54 and a quite high fill factor of 0.54. This suggests that MoO₃-1 is a considerably stable HTL material with unalterable surface chemistry, morphology or electric properties. Figure 2(c) shows the solar photocurrent characteristics of devices comprising fresh or exposed MoO₃-2 HTLs. Three of the samples, **MoO₃-2/Fresh**, **MoO₃-2/Air** and **MoO₃-2/Air & light** exhibit very similar characteristics with $J_{SC}$ around -10.3 mA/cm², $V_{OC}$ of 0.56 and FF of 0.55. Only the photocurrent output of the humidity-exposed sample **MoO₃-2/H₂O** is significantly lower with $J_{SC}$ of -8.9 mA/cm², while neither the FF nor the $V_{OC}$ are affected. The fact that FF and series resistance (indicated by the slope near $V_{OC}$) do not change compared to the other samples, but only the current, indicates that the loss of photocurrent is caused by loss of photoactive area contributions, either by local insulating areas due to corrosion of the ITO electrode interface or by degradation of the adjacent organic semiconductor. Though the current slopes in $V_{OC}$ look quite parallel for ***MoO$_3$-2/H$_2$O*** compared to the other exposure conditions, there is a difference in series resistance with an increase of around 9 %, implying that the bulk or interfacial charge transport is slightly affected.

To differentiate any effects caused by the devices' charge transport ability and potential trapping phenomena, charge transport dynamics were studied via TPC. Complete sets of biased photocurrent transients following 300 µs square laser pulses for devices with exposed HTLs, for all conditions are shown in the Supporting Information with relative (Figure S1) and normalized currents (Figure S2). Selected normalized transient photocurrents are displayed in Figure 3, for devices of exposed PEDOT:PSS HTL (all conditions) and for the humidity-exposed MoO$_3$ HTLs. Comparing the device response of ***PEDOT:PSS/Fresh***, ***PEDOT:PSS/Air***, ***PEDOT:PSS/Air& light*** and ***PEDOT:PSS/H$_2$O*** in Figure 3 (a)-(d), one particular feature can be noticed, that is a current-overshoot of the device at the start of the square light pulse for some samples. For ***PEDOT:PSS/Fresh*** this feature occurs quite weakly and merely for higher positive voltages. For ***PEDOT:PSS/H$_2$O*** on the other hand the effect is present starting from -0.2 V and gets more pronounced the more positive the voltage. ***PEDOT:PSS/Air*** and ***PEDOT:PSS/Air& light*** do not show this behavior. Otherwise the transient characteristics of the devices with exposed PEDOT:PSS are quite similar with rise ($t_{90}$) of $t_{90\ (PEDOT:PSS)}$ = 25 +/- 2 µs and fall times ($t_{10}$) of $t_{10\ (PEDOT:PSS)}$ = 25 +/- 2 µs, respectively. With one exception, PEDOT:PSS/H$_2$O has slightly extended rise and fall times of $t_{90\ (PEDOT:PSS/H2O)}$ = 27 µs and $t_{10\ (PEDOT:PSS/H2O)}$ = 33 µs. In devices with MoO$_3$ HTL current shoots are not observed for any condition. Exemplarily, the photocurrent transients of ***MoO$_3$-1/H$_2$O*** and ***MoO$_3$-2/H$_2$O*** are shown in Figure 3 (e) and (f). The transients of the MoO$_3$ HTL devices behave quite similar for any of the exposure conditions or bias voltage (see also Supplemental Information Figure S2) with rise and fall times of around $t_{90}$ $_{(MoO3-1)}$ = 29 +/- 1 µs and t$_{10\ (MoO3-1)}$ = 30 +/- 1 µs, and of around t$_{90\ (MoO3-2)}$ = 27 +/- 1 µs and t$_{10\ (MoO3-2)}$ = 26 +/- 1 µs, respectively. With one exception, MoO$_3$-2/H$_2$O has slightly extended rise and fall times of t$_{90\ (MoO3-2/H2O)}$ = 30 µs and t$_{10\ (MoO3-2/H2O)}$ = 30 µs. The slight homogeneous increase of charge extraction times when the applied bias is varied from -0.4 V to 0.4 V, is directly assigned to the influence of decreased internal electric field. Lack of heavy or abrupt changes to the rise and fall times and absence of any overshoot behavior are signs for stable, unhindered and well balanced transport in the devices.[45] Especially the biasing is a valuable tool to identify energetic barriers or traps in a device, as the applied voltage depending on polarity and strength helps photogenerated charges to overcome barriers, getting released from traps and swept out of the device.[40, 46] In the case of MoO$_3$-HTL device sets, where divergent behavior could have been expected after the general J-V characteristics, transients show no signs of delayed charges or blocking behaviour. Only ***MoO$_3$-2/H$_2$O*** shows slightly longer rise and fall times, indicating a lower charge transport efficiency of solar cell influenced by the HTL's or interfacial contact resistance. The case of PEDOT:PSS is there somehow more complex. While the air- and air/light exposure shows no signs for blocking of delayed transport either, the fresh and the humidity-exposed film seem to induce a build-up of charges in the device upon switch-on, seen as the overshoot of current at the start of the square laser pulse. This is clearly a consequence of an interface issue near the ITO electrode or near the active layer. The fact that for the fresh one this only happens weakly and only for high positive voltages, reducing the internal electric field of the device, supports the aforementioned theory of an unrelaxed PEDOT:PSS layer. When the bias is strong enough to overcome this eventual layer contact problem, the device transport is normal. For the humidity-exposed PEDOT:PSS layer the consequences seem to be more severe, as the charge build-up already takes effect from negative biases and builds up fast when reducing the device's internal field. This behavior is typical for traps in a device. [40, 46] Merely for negative voltages exceeding -0.3 V the charge transport is unhindered, giving an idea about the strength/depth of the charge traps in these devices. The longer rise and fall times also imply a lower charge transport efficiency of solar cell

Thereby the rising relative magnitude of the overshoot transient peaks to the steady state photocurrent with increasing applied voltage, correlates with trap-facilitated recombination [37]. In this case their origin could be migrated indium or oxidation sites in the organic active layer, or at surface defects (e.g. dangling bonds) of the corroded inorganic ITO. [36]. The build-up of trapped charges in the device may hereby lead to increased recombination rates either directly through trap-assisted recombination routes such as Shockley-Read-Hall recombination, or through space charge mediating increases in bimolecular recombination and/or decreases in charge separation efficiency.[37] When the overshoot transient peaks are not observed in the reversed bias region this indicates that the larger internal electric field in the device inhibits trap-mediated recombination.[46]

XPS and Raman spectroscopy were investigated to understand changes in surface chemistry and local bulk chemistry of HTL films after different environmental exposure, which might explain their impact on the adjacent layers within the solar cell stack. The films' XPS core level spectra of atoms relevant to the respective material are shown in Figure 4. The *S 2p* and *O 1s* core level spectra in **PEDOT:PSS** HTLs for different exposure conditions are presented in Figure 4(a) and (b), respectively. In the *S 2p* signal of PEDOT:PSS (Figure 4a), two weak low-binding energy peaks appear at approximately 165.1 eV and 163.9 eV, corresponding to the spin-split components of the sulfur atoms in the PEDOT, while the stronger high-binding energy peaks at 169.0 eV and 167.8 eV correspond to the sulfur atoms in PSS.[47-48] The intensity relation between PEDOT and PSS contributions is typical, because the PEDOT:PSS particles and films thereof are covered with PSS, therefore diminishing the PEDOT signal. However, no clear difference is seen for different exposure scenarios. Figure 4(b) shows the XPS spectra of *O 1s* core level, a main peak at 531.5 eV that originates from $SO_3^{2-}$ acid groups of PSS, and a weaker shoulder at 533.0 eV due to C-O bonds of PEDOT.[49] Also these spectra exhibit no clear changes on surface chemistry of PEDOT:PSS HTLs. It should be mentioned though that water eventually adsorbed by PSS in humidity-exposed films might have been released in the XPS's high vacuum chamber before measurement was initialized. Figure4 (c) and (e) show the *Mo 3d* doublet core level spectra of ***MoO₃-1*** and ***MoO₃-2*** films. Both materials demonstrate comparable results for fresh films and for air and air with illumination exposures. The doublet peaks show a slight asymmetric line shape, suggesting the presence of mixed oxidation states of Mo, which slightly shift their energy. The major contribution comes from $Mo^{6+}$ oxidation states with the *3d3/2* peak at 236.0 eV and *3d5/2* at 232.8 eV, while the minor underlying peaks centered at 234.7 eV and 231.7 eV are assigned to the 3d orbital doublet of $Mo^{5+}$.[50-51] In both, $MoO_3$-1 and $MoO_3$-2 films, the intensity of the peaks belonging to $Mo^{5+}$ oxidation state get stronger after humid-air exposure, visible a tail of the peaks on the low-energy side. Since for preparation of both materials hydrolysis reactions with a molybdenum precursor are involved, it is suggested absorbed humidity triggered according chemical reactions with Mo atoms in the film, leading to this shift. Figure 4(d) and (f) depict the *O 1s* core level spectra for $MoO_3$-1 and $MoO_3$-2 HTL after exposure. Again here fresh, air- and illuminated air-exposed samples are very similar, showing a main peak at 530.9 eV assigned to the $O^{2-}$ in the oxide [52] and a weak second component at 532.4 eV attributed to hydroxyl groups. [52] With humidity-exposure a significant increase of surface hydroxyl groups is observed and additionally a further signal appears at 533.4 eV, which can be assigned to adsorbed molecular water.[53-54] This confirms an ongoing hydrolysis and condensation reaction of molybdenum oxide in the film upon humidity exposure. However, none of the surface chemical observations can explain the behavior of the films in the device.

Raman spectroscopy provides more detailed information related to the chemical properties of the bulk and is not recorded under vacuum, allowing the exposed films to maintain their chemical surface composition during the measurement. In the present experiment, spatially resolved Raman measurements were performed which allow a point-to-point analysis of the film's chemistry, completing the highly accurate surface sensitive but area-integrated information of XPS. Figure 5 a-c present the averaged spectra of the area scans for all exposition scenarios for PEDOT:PSS, $MoO_3$-1 and $MoO_3$-2, respectively. The complete sets of the corresponding spatially resolved Raman spectra sequences obtained by Raman microscopy scans for all exposure conditions can be found in the Supplemental Information in Figure S3. As the humidity-exposed films seem to exhibit the most relevant changes, these are shown exemplarily for each of the materials in Figure 5 d-f. Figure 5(a) shows the average spectra of point-to-point Raman analysis of **PEDOT:PSS** HTLs after different environmental exposures (see from Figure5 (s) for humid air exposure and Figure S3 in supplement for fresh, air, and air with illumination exposures). All spectra of PEDOT:PSS HTLs after different environmental exposures show nearly identical bands, suggesting no clear chemical composition changes on PEDOT:PSS film after exposure. Those curves exhibit the main peak at 1442 $cm^{-1}$, corresponding to the C=C symmetrical stretching vibration of PEDOT and the band near 1370 $cm^{-1}$ is associated with the C-C stretching. [55] The peaks at 1508 and 1570 $cm^{-1}$ have been associated with the C=C asymmetric stretching vibrations that correspond to thiophene rings in the middle and at the end of the chains, respectively.[56] The band located at 1540 $cm^{-1}$ is related to the splitting of these asymmetrical stretching vibrations.[57-58] All the other bands observed at 440, 578, 701, 854, 991, 1097, 1367, 1255, 1367, 1505, 1539 and 1569 $cm^{-1}$ are also associated with PEDOT on the basis of the spectrum of PEDOT.[55-56] Figure5 (d) demonstrates spatially resolved Raman spectra from the PEDOT: PSS HTL after humid air exposure. The spectra present equal chemical composition across scanning area, merely varying by intensities with film thickness, which corresponds to the bubble feature displayed in the corresponding optical microscopy picture (Fig. 5d inset). These bubbles were exclusively found in humidity-exposed PEDOT:PSS films. It is suggested that the bubbles form as a consequence of film swelling upon water uptake by the hygroscopic polymer PSS. This means that the water exposure does not induce irreversible chemical changes in the PEDOT:PSS layer, but merely reversible water uptake, with eventual effect on morphology.

The average spectra of *MoO$_3$-1* HTLs for freshly prepared, air, air with illumination and humid air exposures are shown Figure 5 (b). These spectra present identical features for all different conditions, indicating no obvious chemical composition transformation on the bulk MoO$_3$-1 HTL films. The broad band features in the spectral are consistent with the previous literatures. [59-61] Thereby the Raman band usually appearing in the 500–1000 cm$^{-1}$ and 200–400 cm$^{-1}$ regions stands for the stretching (ν) and bending (δ) vibration of the structure of MoO$_3$.[60] Here, the broad Raman bands at about 640, 850, 950, and 989 cm$^{-1}$, are explained by a wide distribution of different Mo-O bond length. [59] Bands in the range of 920–1000 cm$^{-1}$ are associated with stretching modes of the Mo=O terminal. [62] Looking at the spatially resolved spectra of humidity exposed MoO$_3$-1 (Figure 5e), all spectra exhibit high homogeneity of peaks and intensities across the scanning area, implying a great chemical stability of MoO$_3$-1 HTL. The microscopic pictures given in the inset of the diagram show a widely continuous and homogeneous film. Figure 5(c) demonstrates average Raman spectra from the four different conditions of *MoO$_3$-2* HTL films. The spectra of fresh MoO$_3$ HTL and after air and air with illumination exposures show identical Raman bands, suggesting the considerably comparable chemical composition in those films. The peaks at 152, 280, 660, 819, and 995 cm$^{-1}$ are associated with the thermodynamically stable α-MoO$_3$ crystalline phase. [63] The vibration modes, which appear in the frequency ranges of 1000–600 cm$^{-1}$ and 600–200 cm$^{-1}$, are corresponding to the stretching and deformation modes, respectively. [64-65] The Raman peaks at 994 and 819 cm$^{-1}$ are the stretching vibration of the terminal double bonds (Mo=O). [66] The wave number range, between 150 and 400 cm$^{-1}$, are associated to the Mo-O modes type scissor, wagging, twist and rotational/translational rigid MoO$_4$ chain mode. [66] Figure 5(f) shows the sequence of spatially resolved Raman spectra of MoO$_3$-2 HTL after humid air exposure. Unlike the PEDOT:PSS and MoO$_3$-1 HTLs after humid air exposure, which present spatially homogeneous chemical composition, MoO$_3$-2 HTLs present considerably inhomogeneous spectra across the scanning area after humid air exposure. These inhomogeneities are not limited to thickness related intensity variations but also different new peaks appearing around 900–1000 cm$^{-1}$ in some local spectra. These additional peaks at 691, 901, 912, and 978 cm$^{-1}$ can be correlated to an h-MoO$_3$ crystalline phase, which has been previously reported. [67-68] The peaks in the range of 880–987 cm$^{-1}$ are due to Mo=O bond and the band at 691 cm$^{-1}$ is corresponding to O–Mo–O vibrations. Most spectra of humidity-exposed MoO$_3$-2 show the peaks of α-MoO$_3$, which is the same phase as seen for all other exposure conditions, but some local spectra display mixed peaks of α-MoO$_3$ and h-MoO$_3$ in one spectrum, implying that water induces a phase transition in the MoO$_3$-2 HTL from α-MoO$_3$ to h-MoO$_3$. From the optical microscopy image, which is shown in the inset Figure 5(f), the film exhibits a granular structure, which is also present in freshly prepared MoO$_3$-2 HTLs and after air and air/illumination exposure. However, upon humidity exposure, these particles seem to swell by absorption of water from around 50 nm for other conditions to about 200 nm. This swelling and also the different MoO$_3$-phase alters the grain boundaries and charge transport properties of the film and thus may lead to lower conductivity of the MoO$_3$-2 HTL film.

To investigate how altered surface and bulk properties of environmentally exposed HTLs and their local homogeneities affect OPV devices they have been integrated in, microscale laser-beam-induced photocurrent mapping was conducted. Figure 6 shows relative photocurrent density (J) maps of P3HT:PCBM solar cells with PEDOT:PSS, MoO$_3$-1 and MoO$_3$-2 HTLs after different environmental exposures with a scanning area of 1500 μm x 4000 μm. On the given length scale, the photocurrent map of **PEDOT:PSS/Fresh** exhibits only subtle wave-like fluctuations across the area of around 5% of highest photocurrent amplitude, except for some considerably lower current near the pixel edges. Further, some very defined spots of 100 μm with slightly lower performance are visible. In comparison, both **PEDOT:PSS/Air** and **PEDOT:PSS/Air & light**, show slightly more homogeneous photocurrent distribution as **PEDOT:PSS/Fresh**, with similar current gradient, but without visible pattern and without mentioned edge effects. For humidity-exposed PEDOT:PSS this changes drastically. **PEDOT:PSS/$H_2O$** shows a highly inhomogeneous photocurrent output, with a wave-like pattern across the surface like **PEDOT:PSS/Fresh**, wherein only 20% of the output area shows unchanged high performance, while for about 2/3 of the area photocurrent output is reduced by 15%, for about 1/10 by 25% photocurrent and output is basically negligible for the remaining area. The fact that air exposure of the HTL neither with nor without illumination has an effect on homogeneity of photocurrent distribution, indicates that PEDOT:PSS itself is not affected by the treatment and also does not trigger defects in the device later. The localized lower photocurrent output spots, which are visible in all the devices with PEDOT:PSS HTLs are assigned to the aggregation of colloidal particles, which has been proven in a previous report.[27] and do not show this edge effects due to the longer relaxation time (during exposure) before device fabrication. The edge effects visible for the fresh device may be caused by aforementioned non-equilibrium interface between ITO and PEDOT:PSS, which affects the film most near the crossover between the ITO pixel patch and the glass substrate. The humidity-exposed device shows the worst performance along the pixel edges for a similar reason. In this case, the moisture uptake makes the PEDOT:PSS film expand. As the film is confined to the substrate dimensions, this leads to folding of the film, thereby creating wave-like current fluctuations. At the crossover between ITO and glass this may lead to partial delamination and thus contact loss. Devices with $MoO_3$-1 HTLs show remarkably comparable homogenious photocurrent distributions for all different environmental exposures **$MoO_3$-1/Fresh**, **$MoO_3$-1/Air**,

*MoO$_3$-1/Air & light* and *MoO$_3$-1/H$_2$O*, merely with a mild smooth photocurrent gradient of less than 5% across the pixel area. This means that the subtle changes in surface chemistry of *MoO$_3$-1/H$_2$O*, which have been observed in XPS, are negligible for the photocurrent homogeneity. This is slightly different for solar cells with MoO$_3$-2 HTLs. They exhibit quite identical homogeneous photocurrent maps for three of the conditions *MoO$_3$-2/Fresh*, *MoO$_3$-2/Air* and *MoO$_3$-2/Air & light*, with a smooth photocurrent gradient of less than 5%, but compared to MoO$_3$-1, with a smaller area fraction of highest output. The homogeneous spatial photocurrent is consistent with the highly homogeneous surface and bulk chemical composition seen from XPS and Raman spectroscopic spectra. In contrast, *MoO$_3$-2/H$_2$O* demonstrates slight inhomogeneities in photocurrent, at the pixel edges, where photocurrent output is decreased down to 85%. From this outcome three possible scenarios can be considered following the absorption of water vapor by the film: swelling of MoO$_3$-2 particles causing lower conductivity of the film (mentioned earlier in context of microscopy), loss of inter-particle/interfacial contact from swelling/shrinking of the particles or from potential ITO corrosion by the remaining acidity in the film, diminishing the conductivity of the actual electrode.

To gain a deeper understanding of the roots for local photocurrent losses in the devices, point-to-point J-V analyses across pixel sections have been conducted, to investigate their device physics. Different from the integrated J-V performance measured under simulated solar conditions (AM1.5G), local photocurrents were measured at 532 nm wavelength, which is at the spectral response maximum of the P3HT:PCBM active layer blend. Figure 7 shows the local photo-J-V curves of devices with the different HTLs after humidity-exposure, recorded along a central scanning line along the long-side of device pixel. These samples have been selected here because merely humidity-exposure displays significant features in the local photocurrent scans, while other conditions *fresh*, *air* and *air&light* are characterized by extraordinary conformity. Complete sets of according scans for all studied conditions of the three materials are presented in Figure S4 of the Supporting Information. For better visibility of any performance variance between local J-V characteristics across a device area, the curves have been color-coded (analogous to the relative photocurrent of the maps in Figure 6) in order of their achieved short-circuit current.

Figure 7(a) shows the point-to-point analysis J-V curves of **PEDOT:PSS/H$_2$O**. All characteristics show consistent slopes at the intersection with V$_{OC}$ and good rectangularity with slight deviation of FF between 0.45 and 0.48, indicating that neither significant series resistance nor interfacial charge transport are affected by humidity in PEDOT:PSS or reactions thereof with adjacent layers. Further, the V$_{OC}$ fluctuates only minorly between 0.40 V to 0.38 V across the device area and the curves exhibit unchanged parallel slopes at the intersection with the current-density axis (J$_{SC}$ region), suspending leakage currents as reason for decreased photocurrent, as the parallel resistance is stable. This is in good agreement with the results from integral device characterization. Therefor the origin for the fluctuating photocurrent across the device area, with J$_{SC}$ varying by 28 % between -0.18 mA/cm$^2$ at optimum performance and 0.13 mA/cm$^2$, is loss of active area. As not leakage, barriers or charge transfer limitations are obvious, photocurrent loss is most likely caused by either local degradation of the active layer by reaction with absorbed water from PSS or by local loss of contact area at the ITO/HTL or HTL/active layer interface. The latter could arise from delamination of the swelling/shrinking PEDOT:PSS layer during water uptake/release during/after device fabrication. Another reason could be failing interfacial wetting during device fabrication upon changes of the HTL's surface chemistry.

Figure 7(b) shows the local J-V characteristics of **MoO$_3$-1/H$_2$O**. Here, considerably uniform J-V curves are visible, which are very similar to the ones of **MoO$_3$-1/Fresh, MoO$_3$-1/Air, MoO$_3$-1/Air & light** shown in the Supporting Information (Figure S4). **MoO$_3$-1/H$_2$O** exhibits exclusively constant V$_{OC}$ of 0.4V, J$_{SC}$ of -0.18 mA/cm$^2$ and FF of 0.57 across the device area. This is in good agreement with their integral device performance and the high homogeneity shown in the photocurrent maps. The minor deviation of photocurrent (around 5% compared to maximum) across the device area is suspected to originate from regular morphological inhomogeneities of the devices [27] and not related to the treatment. Figure 7(c) demonstrates the local variations of J-V curves of **MoO$_3$-2/H$_2$O**. Different to the previous MoO$_3$-1, here fluctuations across the device area are visible. The curves exhibit consistent parallel slopes at the intersection with J$_{SC}$ and with V$_{OC}$ and constant rectangularity with a FF of 0.55, indicating that neither parallel resistance nor series resistance of interfacial charge transfer are affected by humidity from the HTL. In addition, the variation of V$_{OC}$ between 0.39 V and 0.40 V is negligible. Merely the fluctuation of photocurrent between -0.18 mA/cm$^2$ at maximum and lowest of -0.16 mA/cm$^2$ is significant. Thereby this fluctuation of 11% is less than for **PEDOT:PSS/H$_2$O** but considerably stronger than for **MoO$_3$-1/H$_2$O** . As similar for **PEDOT:PSS/H$_2$O** also here it is obvious that this effect is caused by a loss of active area, as suggested either from active layer degradation [69] or by loss of interfacial contact of the HTL with adjacent layers. This would not be unlikely because particle swelling of MoO$_3$-2 upon water-exposure was observed earlier. Thereby it can be speculated that the difference in intensity of this device "degradation" effect between **PEDOT:PSS/H$_2$O** and **MoO$_3$-2/H$_2$O** is related to the amount absorbed water and thus volume fluctuation of the layer.

For a complete overview with greater detail on all HTLs and conditions, the solar cell parameters were extracted/calculated from the respective photo-J-V characteristics in Figure 7 and in the Supplemental Information Figure S4. They are plotted in Figure 8 for J$_{SC}$ (a), V$_{OC}$ (b), FF (c) and R$_S$ (d). As expected from photocurrent maps and their local device-physical analysis, devices with MoO$_3$-1 HTL exhibit clearly the highest spatial homogeneity and stability regarding electrical performance independent of different environmental exposures, with very narrow distribution of less than 5% deviation of the values for J$_{SC}$, V$_{OC}$, FF and R$_S$. In contrast, the devices with PEDOT:PSS HTL present most strongly fluctuating performance and inhomogeneity of local output of $J_{SC}$, $V_{OC}$, FF and $R_S$ for different environmental exposures. In the worst case, which is humidity exposure reflected by **PEDOT:PSS/H₂O**, the deviation is up to 28% for $J_{SC}$ photocurrent and around 6% for its $V_{OC}$, FF and $R_S$.

For **PEDOT:PSS/Fresh**, **PEDOT:PSS/Air**, **PEDOT:PSS/Air & light**, all parameters show a more narrow and very similar distribution, with deviations as high as around 8 % for $J_{SC}$ and between 5 and 6 % for $V_{OC}$, FF and $R_S$, indicating that air and illumination (equivalent of 10% of AM1.5G) do not cause obviously localized surface and electric property changes of the HTL. **MoO₃-2/Fresh**, **MoO₃-2/Air**, and **MoO₃-2/Air & light** present all a quite narrow distribution of solar cell parameters with deviations around 5 % for $J_{SC}$ and less than 3 % for $V_{OC}$, FF and $R_S$. **MoO₃-2/H₂O** on the other hand, exhibits a wider distribution of $J_{SC}$ with a deviation of 10% and less than 4% for $V_{OC}$, FF and $R_S$, however, no inhomogeneity effect on $R_S$ in comparison with other environmental exposures of MoO₃-2 HTL. These results clearly show that the values of PEDOT:PSS HTLs are spreading towards lower $J_{SC}$, slightly lower $V_{OC}$, slightly lower FF and higher series resistance upon HTL contact with humidity. For the MoO₃ HTLs, MoO₃-1 shows no scattering of the values independent of the exposure condition, while for MoO₃-2, clearly a subtle shift in $J_{SC}$ and $V_{OC}$ is noted but with relatively not much increased scattering. Since merely representative data have been selected for display in the main text for the sake of better visibility, statistical data on the devices' variations in performance are presented in the Supplemental Information Figure S5 for completion.

**Discussion**

The presented results show clearly that PEDOT:PSS as HTL induces the most unstable, fluctuating and inhomogeneous performance in standard OPV devices, compared to the two MoO₃ HTLs. Thereby humidity-exposure of PEDOT:PSS before device completion strongest affects electrical performance, despite the fact the XPS and Raman spectra show no change in surface and bulk chemistry of the layer, suspending ITO corrosion as a potential reason and strengthening the argument of layer delamination by consecutive swelling and shrinking of the film upon water uptake or release. A significant current-overshoot and extended charge extraction times seen in the transients confirm less efficient charge transport near the electrode, leading to an accumulation of charges at high light intensities/ current densities. A highly inhomogeneous spatial photocurrent output with a wave-like pattern across the surface supports the idea of interfacial contact issues from partial delamination of PEDOT :PSS from ITO or/and active layer. This in consequnce leads to a general lower photocurrent in the device. The good rectangularity with constant FF, minor $V_{OC}$ fluctuation and stable parallel resistance from point-to-point J-V analysis, showing no significant changes of the cell parameters exept for photocurrent and series resistance, again confirms this assumption. Another condition which surprisingly causes performance changes, is the immediate use of a freshly deposited and dried PEDOT:PSS layer, leading to equal $J_{SC}$ and $V_{OC}$ of, but considerably lower fill factor and an associated higher series resistance of the device, compared to the ones containing air or illumination exposed PEDOT:PSS. The reason is suspected to be from limited interfacial charge transport between the PEDOT:PSS and neighboring layers due to the unrelaxed PEDOT:PSS film. This argument is supported by the fact that any form of short storage before device completion improves the device and further by the observation of a weak current overshoot in the transients at high positive bias and subtle wave-like fluctuations in the spatial photocurrent. Dry air or light exposure of the PEDOT:PSS does not lead to significant changes of the actual layer, indicating that PEDOT:PSS itself is not affected by the treatment, nor to changes of the device, confirming that also no defects are triggered later in the device. ***MoO₃-1*** demonstrates the most stable electrical device performance among the tested materials, with unaltered $J_{SC}$, $V_{OC}$, high fill factor and comparable homogeneous spatial photocurrent distributions for fresh, air, air with illumination and humidity exposures. This reflects also in stable charge transport dynamics seen in the transients and bulk chemical stability seen from Raman spectra. The surface chemistry shows a significant increase of surface hydroxyl groups upon humidity exposure, visible from stronger $Mo^{5+}$ contribution in XPS, which seems not to affect the film properties or the device performance. ***MoO$_3$-2*** is reasonably stable for fresh films and after exposure to air or illumination, exhibiting comparable output of all solar cell parameters, $J_{SC}$, $V_{OC}$, FF, and $R_S$, as well as highly homogeneous spatial photocurrent output. However, upon humidity exposure, according devices show lower photocurrent output, obviously induced by a slightly increased series resistance, while other parameters are barely affected. Together with the microscopic observation of MoO$_3$-2 particle swelling/shrinking upon water uptake/release, this can be explained by decreased interparticle-contact, detachment from the ITO surface or increased resistance of the actual swollen particle due to the hydrated (thus insulating) outer shell. A slightly longer charge extraction time in the photocurrent transients but missing overshoot, supports this theory and confirms that the effect on charge transport efficiency is not as strong as for PEDOT:PSS. This is further supported by a homogeneously slightly increased series resistance visible from local variations of J-V curves. A subtle decrease in photocurrent along the pixel edge suggests detachment of the particles at the transition between glass and ITO on the substrate, thus similar but not the same phenomenon as observed for humidity-exposed PEDOT:PSS. The observed phase transition from α-MoO$_3$ to h-MoO$_3$ in the Raman spectra is not relevant to lower charge transport efficiency due to the actual higher conductivity of h-MoO$_3$ compared to α-MoO$_3$. [70] However, it cannot be excluded that the recrystallization changed interparticle or particle/ITO contact.

CONCLUSIONS

We compared the surface and device inhomogeneity of P3HT: PCBM bulk heterojunction PV cells influenced by three different solution-processed colloidal HTLs with various environmental exposures, one PEDOT: PSS, one continuous $MoO_3$ and one nanoparticle $MoO_3$ film. HTLs are exposed to the air in the dark, air with equivalent 10% AM1.5G solar illumination, and humidity in the dark, and compare them with fresh samples, respectively, to gain the information about potentially occurring degradation mechanisms influenced by HTLs with different environmental effects. Standard OPV device investigation is accompanied by the comparison of surface and bulk chemistry of the HTLs, along with the devices' charge transport dynamics, spatial photocurrent distribution and local device physics. By this exposure procedure limited to the actual HTL, the environmental effects caused directly on the HTL and potential late effects within the device structure could be distinguished without affecting the entire device stack, different to studies with exposure of the complete device. Here we could clearly show that humidity exposure of PEDOT:PSS does not actually lead to device failure by the suspected ITO corrosion or humidity diffusion into the active layer with associated degradation, but instead merely signs for partial interfacial contact loss were found. These are caused by the consecutive swelling and shrinking of the layer on top of the ITO substrate and after incorporation within the device stack. Also fresh PEDOT:PSS layers which are immediately incorporated into devices present interfacial resistance, but, in that case because the layer is in a completely dry rigid non-equilibrium state at use. Any form of relaxation time leads to improvement. The two investigated solution processed molybdenum oxide films show extremely stable unchanging electrical behavior and homogeneity for any condition, except for humidity exposure of the particle layer. In that case humidity treatment leads to swelling of the particles, similar to PEDOT:PSS, but due to the thin non-continuous layer, the effect leads to generally higher series resistance and thus lower current by either detachment of particles from the other layers or increased resistance of the particles themselves by the insulating hydrate shell. In no condition and for no material any material degradation was found for the HTL itself or the device they were incorporated in. The observation of decreased device performance merely by the loss of interfacial contact upon water-uptake induced swelling is a completely new view on the matter.

ASSOCIATED CONTENT

**Supporting Information**

The Supporting Information contains complete sets of relative and normalized transient photocurrents, spatially resolved sequences of Raman spectra, statistical data on the solar cell performance and sets of local photo-J-V curves across the device for different hole-transportation layer materials after various environmental exposures.

AUTHOR INFORMATION

**Corresponding Author**

*Email: hchien@tugraz.at

**Notes**

The authors declare no competing financial interest.

ACKNOWLEDGMENT

H.-T. C. and B.F. are grateful to the Austrian Science Fund (FWF) for financial support (Project No. P 26066)

(71) Wang, W.; Guo, S.; Herzig, E. M.; Sarkar, K.; Schindler, M.; Magerl, D.; Philipp, M.; Perlich, J.; Müller-Buschbaum, P. Investigation of morphological degradation of P3HT:PCBM bulk heterojunction films exposed to long-term host solvent vapor. *J. Mater. Chem. A* **2016**, *4*, 3743–3753

FIGURES

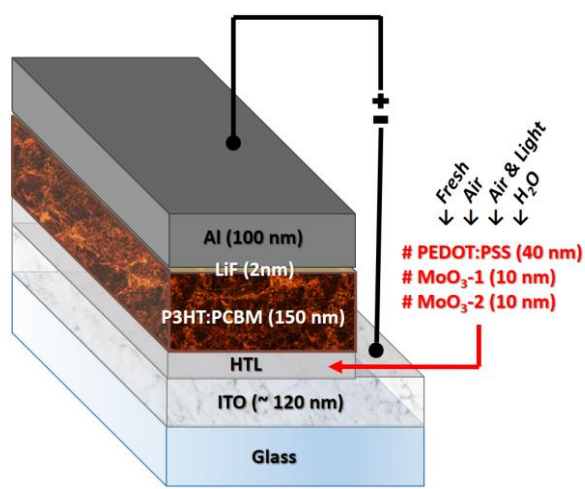

**Figure 1.** Investigated device architecture with the focus on different hole-transport layers and their environmental exposure conditions.

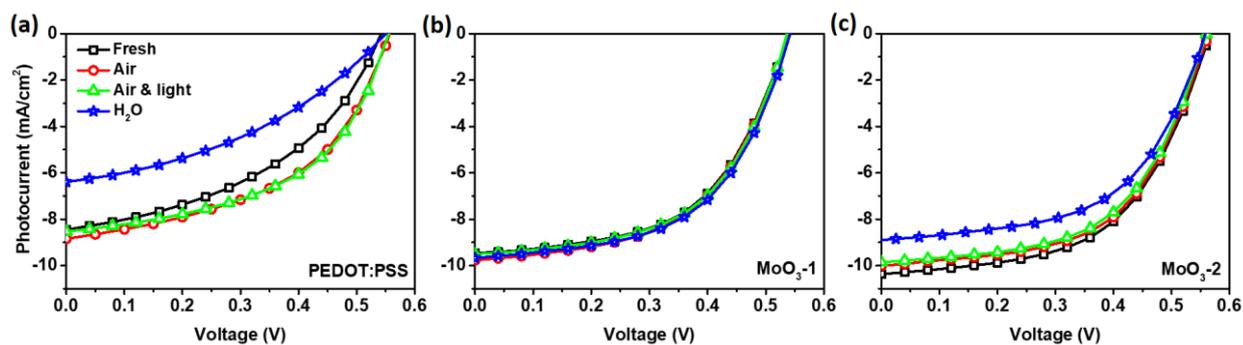

**Figure 2.** Photocurrent characteristics under AM1.5G of devices with fresh prepared HTLs and after different environmental exposures (a) PEDOT:PSS (b) MoO$_3$-1 (c) MoO$_3$-2.

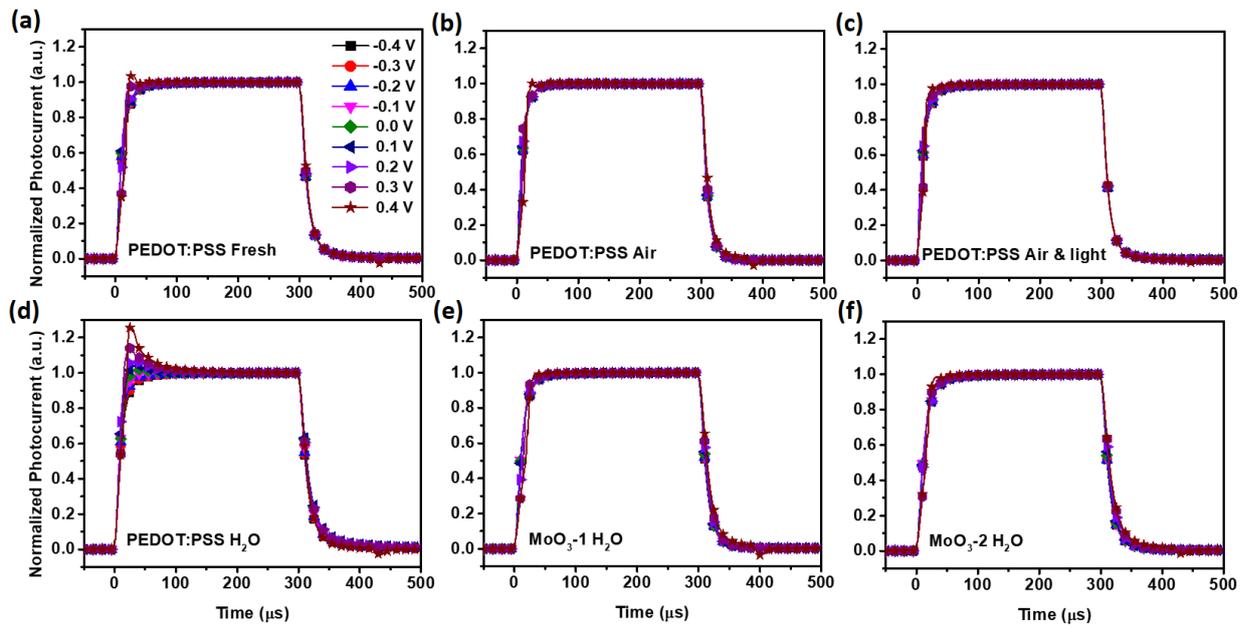

**Figure 3.** Normalized TPC (normalized by the steady state photocurrent at 300 μs) at different applied bias of selected devices with different environmentally exposed HTLs, namely (a) freshly prepared PEDOT:PSS (b) PEDOT:PSS after air exposure in the dark (c) PEDOT:PSS after air with illumination (d) PEDOT:PSS after humid air exposure in the dark (e) MoO$_3$-1 after humid air exposure in the dark (f) MoO$_3$-2 after humid air exposure in the dark.

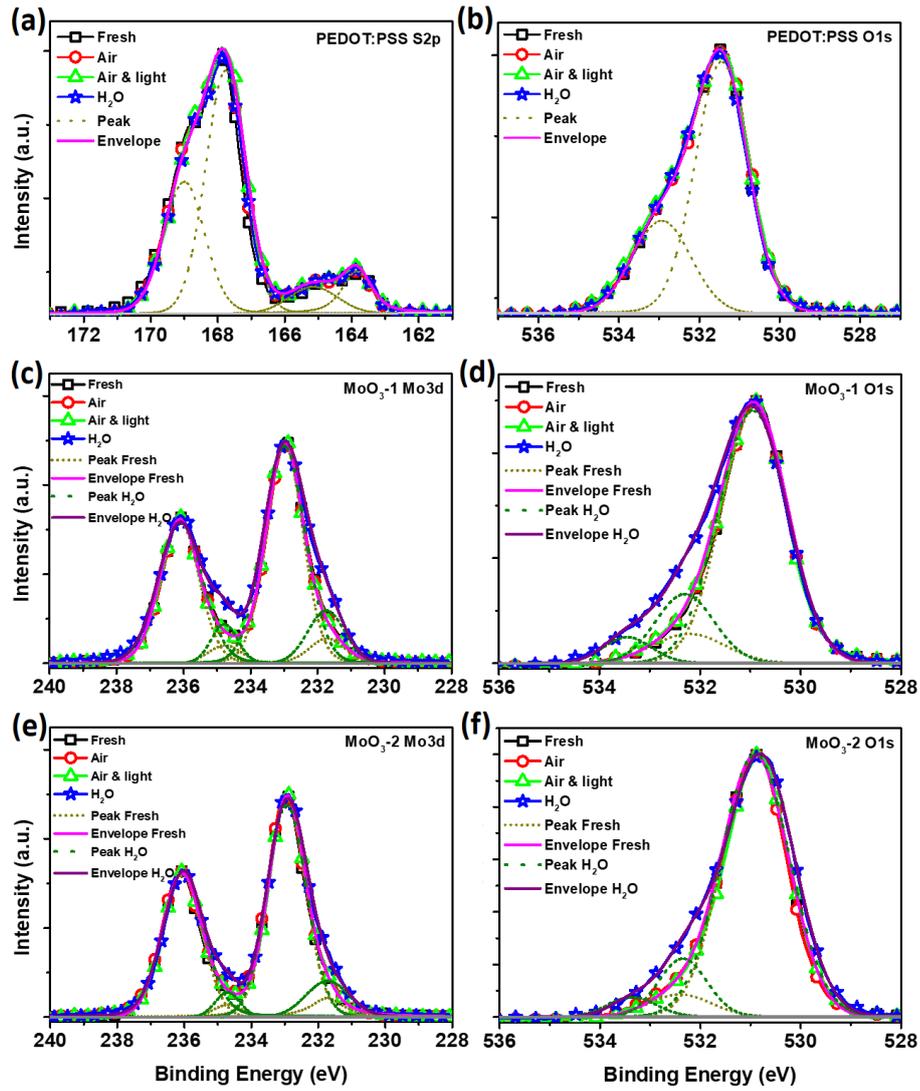

**Figure 4.** XPS spectra of different HTL films after selected environmental exposures: (a) S 2p core level of PEDOT:PSS (b) O 1s core level of PEDOT:PSS (c) Mo 3d core level of MoO$_3$-1 (d) O 1s core level of MoO$_3$-1 (e) Mo 3d core level of MoO$_3$-2 (f) O 1s core level of MoO$_3$-2.

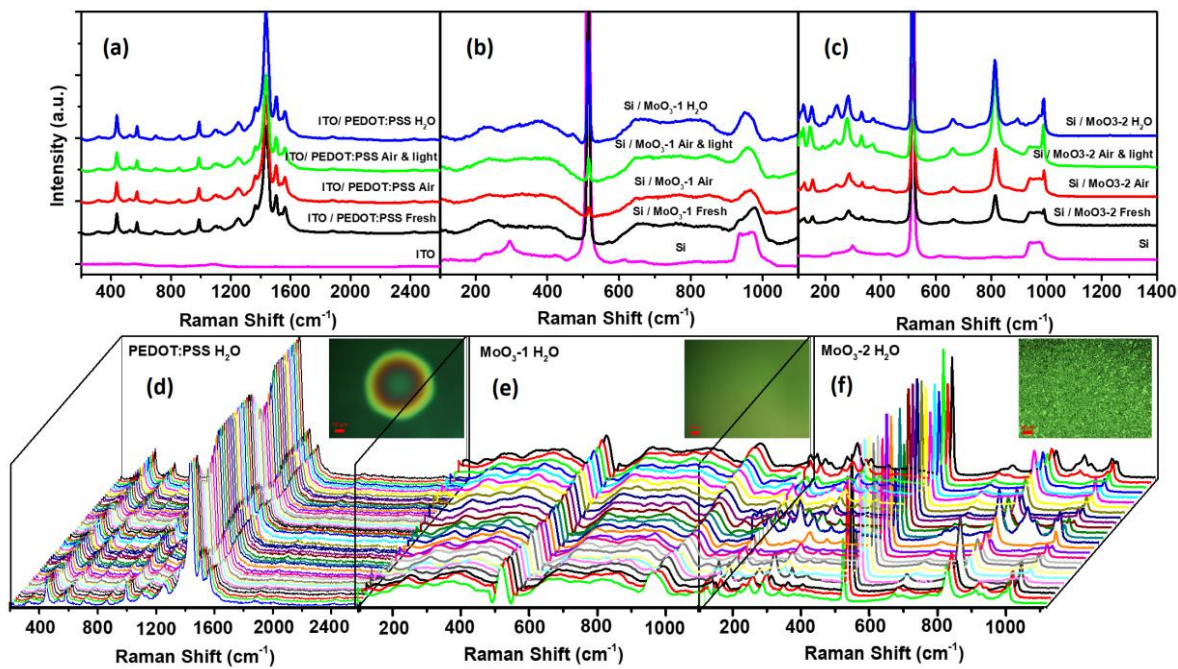

**Figure 5.** Area-averaged spectra of spatially resolved Raman characteristics of PEDOT:PSS (a), $MoO_3$-1 (b) and $MoO_3$-2 (c) films after different exposures scenarios. Spatially resolved sequences of Raman spectra for humidity-exposed PEDOT:PSS (d), $MoO_3$-1 (e) and $MoO_3$-2 (f).

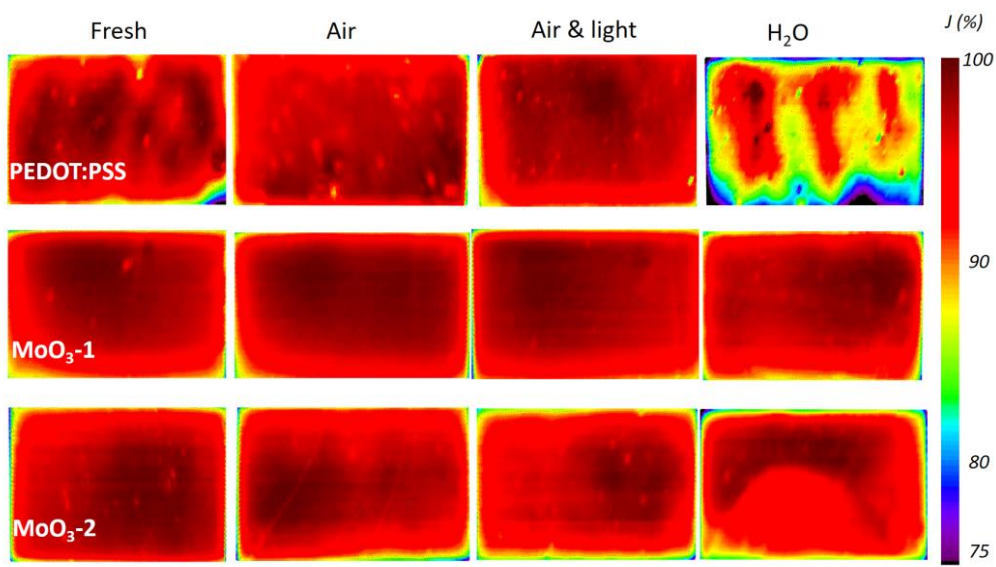

**Figure 6.** Relative photocurrent density maps of P3HT:PCBM solar cells with PEDOT:PSS, MoO$_3$ -1 and MoO$_3$ -2 HTLs after different environmental exposures with a scanning area of 1500 μm x 4000 μm.

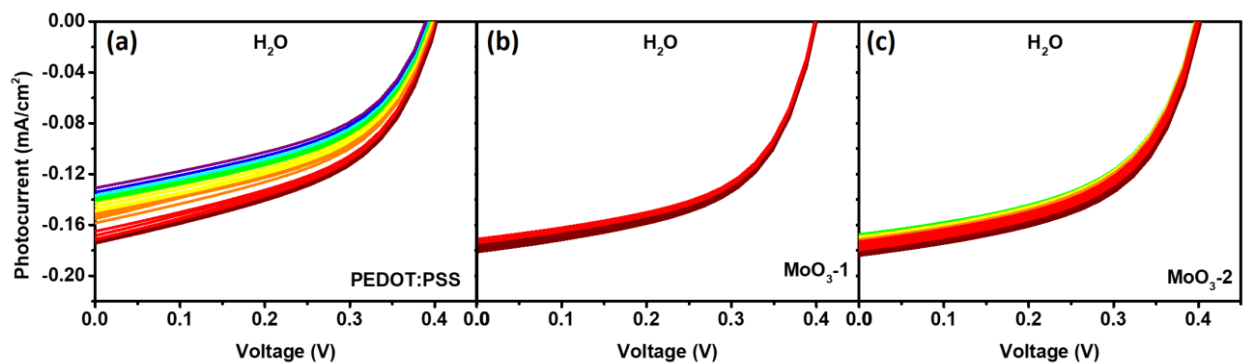

**Figure 7.** Local photo-J-V curves from a central line-scan along the long-side of devices with different HTL materials after humid air in the dark exposure (a) PEDOT:PSS (b) MoO$_3$-1 (c) MoO$_3$-2.

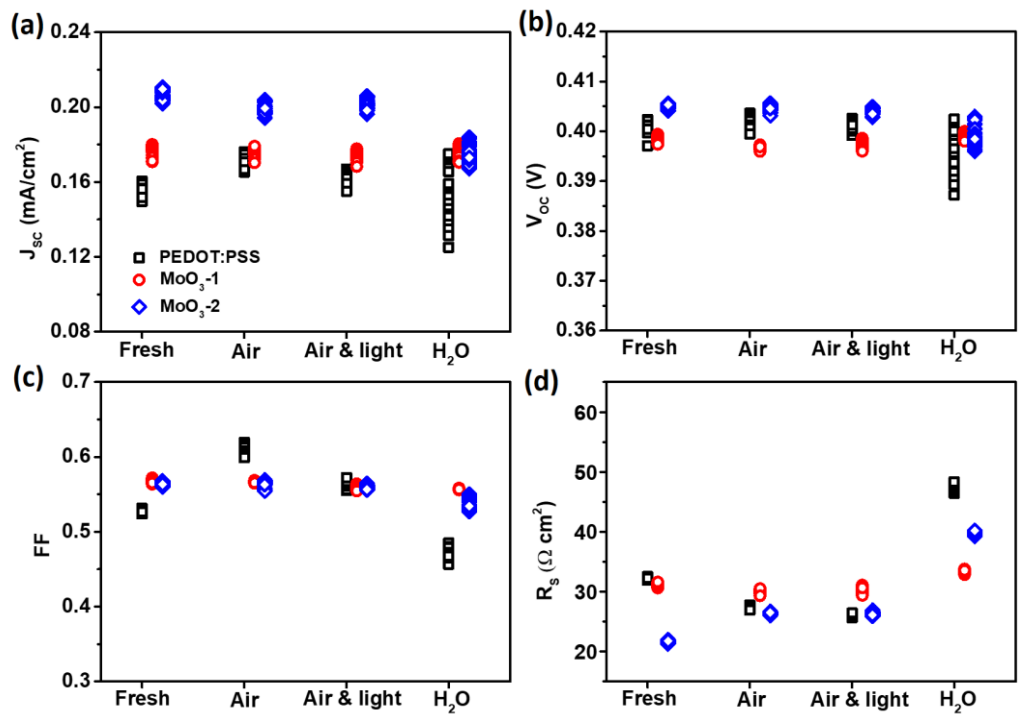

**Figure 8.** Solar cell parameters as extracted/calculated from the local photo-J-V curves recorded at different points along the long-side (4 mm) of a device pixel, showing (a) $J_{SC}$, (b) $V_{OC}$, (c) FF and (d) $R_S$.

SYNOPSIS TOC

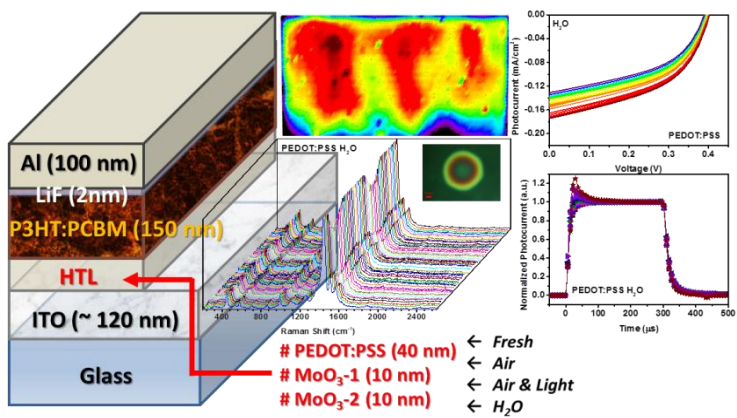